\documentclass[showpacs,groupedaddress,pra,letterpaper,twocolumn]{revtex4}
\usepackage{graphicx}

\input{tcilatex}

\begin{document}

\title{Entanglement reciprocation between atomic qubits and entangled
coherent state}
\author{Ling Zhou, Guo-Hui Yang}
\affiliation{Department of Physics, Dalian University of Technology, Dalian 116024,
P.R.China}

\begin{abstract}
Introducing classical fields, we can transfer entanglement completely from
discrete qubits into entangled coherent state. The entanglement also can be
retrieved from the continuous-variable state of the cavities to the atomic
qubits. Via postselection measure, atomic entangled state and entangled
coherent state can be mutual transformed fully.
\end{abstract}

\maketitle

\section{\protect\bigskip Introduction}

Entanglement transfer between qubits and continuous-variable systems is an
key step in constructing quantum network. There are several proposals to
construct a quantum network \cite{ciracr,kraus,lee,plenio}. In Kraus \cite%
{kraus} recent scheme, quantum network can be established by using photons
to entangle atoms which are located at different nodes for storing quantum
information. Under two-mode squeezed vacuum environment, the entanglement of
two-mode squeezed light is transferred into atoms. Most recently, Lee\cite%
{lee} put forward a proposal reversely, which use continuous-variable state
to store memory. This scheme has its advantages. Because of the infinite
dimensions of their Hilbert space, it is believed that multidimensional
state can store entanglement memory much more than one ebit \cite{serafini}.
In the two ways of quantum network, high efficiency transfer between qubits
and continuous-variable state are both important. Therefore, a lot of
efforts have been devoted to enhance dynamic entanglement transfer efficiency%
\cite{serafini}\cite{paternostro}\cite{zou}\cite{won}\cite{ret}. Serafini 
\cite{serafini}presented strategies to enhance the dynamic entanglement
transfer from continuous-variable to finite-dimensional systems by employing
multiple qubits. Paternostro\cite{paternostro} showed how the quantum
correlations initially present in the driving field play a critical role in
the entanglement transfer from Gaussian state to qubit. Zou \cite{zou}
studied the possibility of entangling two separable and mixed qubits by
local interaction with the two-mode nonclassical field state.

In this paper, we introduce two classical fields to drive two atoms. It is
found that the entanglement of two atoms can be transferred into entangled
coherent state with one hundred percent efficiency. We also can retrieve the
\ entanglement from the continuous -variable state to the atomic qubits with
maximum entanglement. Via postselection measure, atomic entangled state and
entangled coherent state can be mutual transformed completely. The scheme
can be realized in cavity QED system.

\section{The theory and the scheme description\protect\bigskip}

\bigskip We assume that the quantum network is established by using two
entangled atoms to deposit information in their cavities or two
non-entangled atoms to retrive information from the two entangled cavity
fields. We \ consider that two identical atoms interact with two separate
cavities $A$ and $B$ while the two atoms are driven by two classical fields.
In the Hilbert space $\Re $, the Hamiltonian of the total system is

\begin{equation}
H=\sum_{j=1,2}\Omega _{j}(\sigma _{j}+\sigma _{j}^{\dagger })+\lambda
_{j}(a_{j}^{\dagger }\sigma _{j}+a_{j}\sigma _{j}^{\dagger }),  \label{eq.1}
\end{equation}%
where $\Omega _{j}$ is the Rabi frequency of two classical fields, and $%
\lambda _{j}$ is the coupling between the cavities and the atoms. $a_{j}$
and $a_{j}^{\dagger }$ are the annihilation and creation operators for the
two cavities, respectively. $\sigma _{j}=|g\rangle _{jj}\langle e|$ and $%
\sigma _{j}^{\dagger }=|e\rangle _{jj}\langle g|$ are transition operators
of the two atoms where $|e\rangle $ and $|g\rangle $ stand for the excited
and ground states of the atoms. Because usually the classical field is very
stronger than the quantum field, we will take classical field as main part
and take the quantum field as perturbed term. This strategy has been used in 
\cite{solano}. In order to to perform a unitary transformation easily, we
sort the Hamiltonian as 
\begin{equation}
H_{0}=\sum_{j=1,2}\Omega _{j}(\sigma _{j}+\sigma _{j}^{\dagger })
\label{eq.2}
\end{equation}%
\begin{equation}
H_{I}=\sum_{j=1,2}\lambda _{j}(a_{j}^{\dagger }\sigma _{j}+a_{j}\sigma
_{j}^{\dagger })  \label{eq.3}
\end{equation}%
Changing the atomic bare-state basis into dressed-state basis, i.e., 
\begin{equation}
|\pm _{j}\rangle =\frac{1}{\sqrt{2}}(|g_{j}\rangle \pm |e_{j}\rangle ,
\label{eq.4}
\end{equation}%
and performing the unitary transform $U_{0}(t)=e^{-iH_{0}t}$ on $H_{I}$, in
strong driving regime $\Omega _{1}(\Omega _{2})\gg \lambda _{1}(\lambda
_{2}) $, we can realize a rotating-wave approximation and eliminate high
frequency term as it was done in \cite{solano}. After the transformation, in
the new Hilbert space $\Re ^{\prime }$, the effective Hamiltonian is 
\begin{equation}
H_{eff}=\sum_{j=1,2}\frac{\lambda _{j}}{2}(|+_{j}\rangle \langle
+_{j}|-|-_{j}\rangle \langle -_{j}|)(a_{j}^{\dagger }+a_{j}).  \label{eq.5}
\end{equation}%
In next section, we will use the effective Hamiltonian to calculate the
entanglement transfer.

\section{\protect\bigskip Entanglement transfer from qubits to entangled
coherent state}

We assume that initial state of the two cavities are both in vacuum state
while the two atoms are in maximal entangled state, that is, the initial
state of the system is as 
\begin{equation}
|\Psi (0)\rangle =\frac{1}{\sqrt{2}}(|eg\rangle +|ge\rangle )|0,0\rangle
\label{eq.6}
\end{equation}%
Now, we want to deposit the atomic entangled information into the field
state and let the two entangled atoms enter into their cavities separately.
The evolution of the system state under the Hamiltonian Eq.(5) can be
deduced as 
\begin{equation}
|\Psi (t)\rangle =\frac{1}{\sqrt{2}}(|+,+,\alpha _{1},\alpha _{2}\rangle
-|-,-,-\alpha _{1},-\alpha _{2}\rangle ),  \label{eq.7}
\end{equation}%
where $\alpha _{j}=-\frac{i\lambda _{j}t}{2}$. Note that here $|\Psi
(t)\rangle $ in Eq.(7) is not normalized state. We change the atomic basis
into the bare atomic basis $\{|e_{i}\rangle ,|g_{i}\rangle \}$ and inverse
the unitary transformation. Thus, the state of the system in the space $\Re $
can be rewritten as 
\begin{eqnarray}
|\Psi (t)\rangle &=&\frac{1}{2\sqrt{2}}[|gg\rangle (e^{iut}|\alpha
_{1},\alpha _{2}\rangle -e^{-iut}|-\alpha _{1},-\alpha _{2}\rangle ) 
\nonumber \\
&&+|ge\rangle (e^{iut}|\alpha _{1},\alpha _{2}\rangle +e^{-iut}|-\alpha
_{1},-\alpha _{2}\rangle )  \nonumber \\
&&+|eg\rangle (e^{iut}|\alpha _{1},\alpha _{2}\rangle +e^{-iut}|-\alpha
_{1},-\alpha _{2}\rangle )  \nonumber \\
&&+|ee\rangle (e^{iut}|\alpha _{1},\alpha _{2}\rangle -e^{-iut}|-\alpha
_{1},-\alpha _{2}\rangle )],  \label{eq.8}
\end{eqnarray}%
where $u=\Omega _{1}+\Omega _{2}$. The state in Eq.(8) is still not
normalized. When the atoms come out from the two cavities, we use
level-selective ionizing counter and detect the atomic state. If the
internal state of atoms are detected at any of the states $|ee\rangle
,|eg\rangle ,|ge\rangle ,|gg\rangle ,$ the state of the two-cavity fields
are projected into

\begin{equation}
|\Psi (t)\rangle _{f\pm }=\frac{1}{\sqrt{M_{\pm }}}[e^{uti}|\alpha
_{1},\alpha _{2}\rangle \pm e^{-uti}|-\alpha _{1},-\alpha _{2}\rangle ],
\label{9}
\end{equation}%
where%
\begin{equation}
M_{\pm }=2[1\pm \cos 2ut\exp (-2|\alpha _{1}|^{2}-2|\alpha _{2}|^{2})].
\label{10}
\end{equation}%
Here the state in Eq.(9) is normalized. One can see that the atomic
entangled state is completely transferred into the two cavity entangled
coherent state if the state in Eq.(9) state is a maximal entangled one. We
recall that the concurrence of state can be used to estimate the
entanglement for such a state \cite{wang,xgwang}. The concurrence\cite%
{wooters} of a state is defined as $C=\max (0,2\max {\chi _{i}}%
-\sum_{i=1}^{4}\chi _{i})$ where $\chi _{i}$ is the square roots of the
eigenvalues of the matrix $R=\rho (\sigma _{1}^{y}\otimes \sigma
_{2}^{y})\rho ^{\ast }(\sigma _{1}^{y}\otimes \sigma _{2}^{y})$ and $\sigma
_{i}^{y}$ is Pauli matrix. The concurrence of the state generated from our
system (Eq.(9)) is 
\begin{equation}
C=\frac{\sqrt{[(1-\exp (-4|\alpha _{1}|^{2})][(1-\exp (-4|\alpha _{2}|^{2})]}%
}{1\pm \cos 2ut\exp (-2|\alpha _{1}|^{2}-2|\alpha _{2}|^{2})}  \label{eq11}
\end{equation}%
\begin{figure}[h]
\includegraphics*[width=70mm,height=50mm]{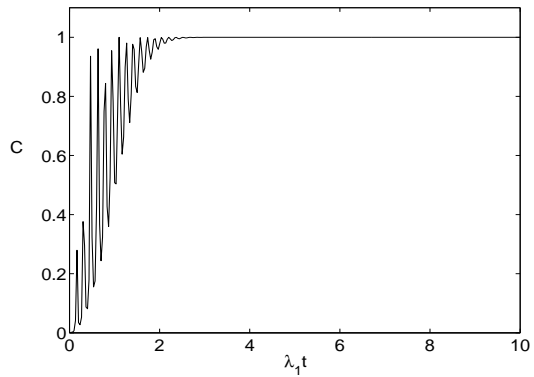}
\caption{The entanglement in two-mode fields state, where $\protect\lambda %
_{1}=\protect\lambda _{2}=1$, $\Omega _{1}=\Omega _{2}=20$.}
\end{figure}
where $+(-)$ correspond to $+(-)$ in Eq.(9). Due to the factor $u$ , the
entanglement of state Eq.(9) is different with the state $\frac{1}{N_{\pm }}%
(|\alpha ,\alpha \rangle \pm |-\alpha ,-\alpha \rangle )$ \cite{xiaoguang}%
\cite{VAN} even when $\alpha _{1}=\alpha _{2}$. In their case, $\frac{1}{%
N_{-}}(|\alpha ,\alpha \rangle -|-\alpha ,-\alpha \rangle )$ is exact one
ebit and its entanglement is always 1 while $\frac{1}{N_{+}}(|\alpha ,\alpha
\rangle +|-\alpha ,-\alpha \rangle )$ is maximum state only when $\alpha
\rightarrow \infty $. In our case, we find there is little difference
between the two state. If we see the state in space $\Re ^{\prime }$ our
state is exact the same with theirs. When we measure entanglement in space $%
\Re $, we need to use the form of Eq.(9). Its Concurrence is show in Fig.1
where we choose negative in Eq.(11) and all the parameters are
dimensionless. Fig.1 show that Concurrence exhibits a rapid oscillation
during a section of starting time which results from the classical field
oscillation. After the time evolution section, Concurrence achieves its
maximum value 1 and the state is a maximum entangled state. This is because
the quantity of $|\alpha _{j}|=\frac{\lambda _{j}t}{2}$ is increased with
time $t$ evolution. If the quantity of $\alpha _{j}$ is large enough, $i.e.,$
$\langle \alpha _{j}|-\alpha _{j}\rangle =\exp (-2|\alpha _{j}|^{2})=0$,
Concurrence in Eq.(11) will be maximum value 1 and the state of Eq.(9) will
be a ebit state. Thus with time evolution, the two-mode field will be a
maximum entangled state so that the entanglement in qubit is completely
transferred into two-mode fields state.

\section{Entanglement retrieval from entangled coherent state to qubits}

We have transferred a maximal entanglement to a two-cavity fields
completely. Next, we will retrieve the deposit entangled information from
two-cavity fields, that is, entanglement transfer from two -cavity fields to
atomic state. We assume the two-cavity fields now are entangled coherent
state with Eq.(9) form which is produced by the above procedure. Now the two
atoms in their ground state $|gg\rangle $ enter into their cavity
separately. So, in space $\Re $ the initial state of the system is 
\begin{equation}
|\Psi (t)\rangle =\frac{1}{\sqrt{M_{\pm }}}[e^{uti}|\alpha _{1},\alpha
_{2}\rangle \pm e^{-uti}|-\alpha _{1},-\alpha _{2}\rangle ]|gg\rangle .
\label{eq.12}
\end{equation}%
where $\alpha _{j}=-\frac{i\lambda _{j}t}{2}$ is a pure complex number which
is exact what we deposit during the process of transfer from qubit to
fields. The state evolution of the system still obey%
\begin{equation}
U(t\prime )=U_{0}^{+}(t^{\prime })e^{-iH_{eff}t^{\prime
}}U_{0}(t)=U_{0}^{+}(t^{\prime }-t)e^{-iH_{eff}t^{\prime }}
\end{equation}%
The evolution will give us state in space $\Re $. We assume $t^{\prime }=t$,
which means we will use the same time to retrieve it back when we use time $%
t $ to deposit it in two-cavity fields. This strategy had been used in \cite%
{lee}. The condition unitary evolution operator now is $U(t\prime
)=e^{-iH_{eff}t^{\prime }}$. Consider the values of $\alpha _{j}$ is pure
complex number, after some calculation, we obtain the state of the system at
evolution time $t^{\prime }$ as

\begin{eqnarray}
|\Psi \rangle &=&[|+,+\rangle (e^{iut}|\alpha _{1}-\frac{i}{2}\lambda
_{1}t^{\prime },\alpha _{2}-\frac{i}{2}\lambda _{2}t^{\prime }\rangle 
\nonumber \\
&&\pm e^{-iut}|-\alpha _{1}-\frac{i}{2}\lambda _{1}t^{\prime },-\alpha _{2}-%
\frac{i}{2}\lambda _{2}t^{\prime }\rangle )  \nonumber \\
&&+|+,-\rangle (e^{iut}|\alpha _{1}-\frac{i}{2}\lambda _{1}t^{\prime
},\alpha _{2}+\frac{i}{2}\lambda _{2}t^{\prime }\rangle  \nonumber \\
&&\pm e^{-iut}|-\alpha _{1}-\frac{i}{2}\lambda _{1}t^{\prime },-\alpha _{2}+%
\frac{i}{2}\lambda _{2}t^{\prime }\rangle )  \nonumber \\
&&+|-,+\rangle (e^{iut}|\alpha _{1}+\frac{i}{2}\lambda _{1}t^{\prime
},\alpha _{2}-\frac{i}{2}\lambda _{2}t^{\prime }\rangle  \nonumber \\
&&\pm e^{-iut}|-\alpha _{1}+\frac{i}{2}\lambda _{1}t^{\prime },-\alpha _{2}-%
\frac{i}{2}\lambda _{2}t^{\prime }\rangle )  \nonumber \\
&&+|-,-\rangle (e^{iut}|\alpha _{1}+\frac{i}{2}\lambda _{1}t^{\prime
},\alpha _{2}+\frac{i}{2}\lambda _{2}t^{\prime }\rangle  \nonumber \\
&&\pm e^{-iut}|-\alpha _{1}+\frac{i}{2}\lambda _{1}t^{\prime },-\alpha _{2}+%
\frac{i}{2}\lambda _{2}t^{\prime }\rangle ).  \label{eq14}
\end{eqnarray}%
Substitute $\alpha _{j}=-\frac{i\lambda _{j}t}{2}$ and $t=t^{\prime }$ into
Eq.(14) and rewrite the state as 
\begin{eqnarray}
|\Psi \rangle &=&|+,+\rangle (e^{iut^{\prime }}|-i\lambda _{1}t^{\prime
},-i\lambda _{2}t^{\prime }\rangle \pm e^{-iut^{\prime }}|0,0\rangle ) \\
&&+|+,-\rangle (e^{iut^{\prime }}|-i\lambda _{1}t^{\prime },0\rangle \pm
e^{-iut^{\prime }}|0,i\lambda _{2}t^{\prime }\rangle )  \nonumber \\
&&+|-,+\rangle (e^{iut^{\prime }}|0,-i\lambda _{2}t^{\prime }\rangle \pm
e^{-iut^{\prime }}|i\lambda _{1}t^{\prime },0\rangle )  \nonumber \\
&&+|-,-\rangle (e^{iut^{\prime }}|0,0\rangle \pm e^{-iut^{\prime }}|i\lambda
_{1}t^{\prime },i\lambda _{2}t^{\prime }\rangle ).  \nonumber
\end{eqnarray}

In order to improve the degree of entanglement transfer to the atoms,
postselection measure on the fields is needed. This procedure is similar to
what we have detected on the atomic state so as to get a pure two--cavity
fields. Here, one can see clearly from Eq.(15) that we will have seven
different kinds of cavity fields.

\subsection{Project with $P=|0,0\rangle \langle 0,0|$}

If we project the two-cavity fields with $P=|0,0\rangle \langle 0,0|$ $($%
means we detect the cavity fields as $|0,0\rangle )$, the measure on fields
results in the atomic state as 
\begin{equation}
|\Psi \rangle =(p_{1}|++\rangle +p_{2}|+-\rangle +p_{3}|-+\rangle
+p_{4}|--\rangle )
\end{equation}%
where%
\begin{eqnarray}
p_{1} &=&e^{-\frac{(\lambda _{1}t^{\prime })^{2}}{2}-\frac{(\lambda
_{2}t^{\prime })^{2}}{2}+iut^{\prime }}\pm e^{-iut^{\prime }}, \\
p_{2} &=&e^{-\frac{(\lambda _{1}t^{\prime })^{2}}{2}+iut^{\prime }}\pm e^{-%
\frac{(\lambda _{2}t^{\prime })^{2}}{2}-iut^{\prime }},  \nonumber \\
p_{3} &=&e^{-\frac{(\lambda _{2}t^{\prime })^{2}}{2}+iut^{\prime }}\pm e^{-%
\frac{(\lambda _{1}t^{\prime })^{2}}{2}-iut^{\prime }},  \nonumber \\
p_{4} &=&e^{iut^{\prime }}\pm e^{-\frac{(\lambda _{1}t^{\prime })^{2}}{2}-%
\frac{(\lambda _{2}t^{\prime })^{2}}{2}-iut^{\prime }}.  \nonumber
\end{eqnarray}%
On the bare atomic basis, the state is 
\begin{equation}
|\Psi \rangle _{a}=N(A_{gg}|gg\rangle +A_{ge}|ge\rangle +A_{eg}|eg\rangle
+A_{ee}|ee\rangle )
\end{equation}%
where 
\begin{eqnarray}
A_{gg} &=&(p_{1}+p_{4}+p_{2}+p_{3}), \\
A_{ge} &=&(p_{1}-p_{4}-p_{2}+p_{3}),  \nonumber \\
A_{eg} &=&(p_{1}-p_{4}+p_{2}-p_{3}),  \nonumber \\
A_{ee} &=&(p_{1}+p_{4}-p_{2}-p_{3}),  \nonumber \\
N &=&\frac{1}{\sqrt{A_{gg}^{2}+A_{ge}^{2}+A_{eg}^{2}+A_{ee}^{2}}}.  \nonumber
\end{eqnarray}%
\begin{figure}[h]
\includegraphics*[width=70mm,height=50mm]{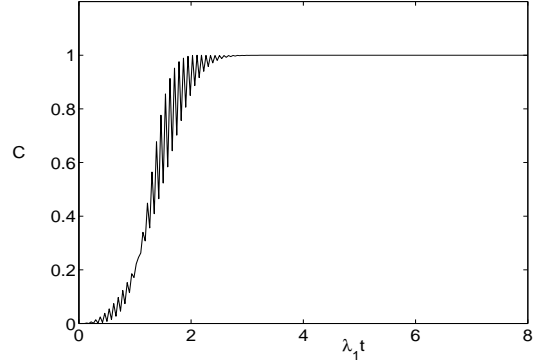}
\caption{The atomic entanglement change with time, corresponding to positive
in Eq.(12). Time here means for different entangled coherent state $\protect%
\alpha _{j}$ we need use corresponding time to retrive it. The parameters
are $\protect\lambda _{1}=\protect\lambda _{2}=1$, $\Omega _{1}=\Omega
_{2}=20$. }
\end{figure}

We still use the concurrence to estimate the entanglement Eq.(18) state. The
Concurrence of the state is 
\begin{equation}
C{\small =2}N^{2}\sqrt{|A_{ee}A_{gg}|^{2}+|A_{eg}A_{ge}|^{2}-A}  \label{eq18}
\end{equation}%
with $A=A_{ee}A_{gg}A_{ge}^{\ast }A_{eg}^{\ast }+A_{ee}^{\ast }A_{gg}^{\ast
}A_{ge}A_{eg}$.

The atomic entanglement is shown in Fig.2. One still can see the oscillation
resulting from classical field $\Omega _{1}$ and $\Omega _{2}$. Excepting
short time section, the entanglement will be maximum value when time $t$ is
longer. We can understand it from the analytic expression Eqs. (17)-(19). If 
$\lambda _{1}=\lambda _{2}$ and large $t$, $p_{2}=p_{3}=0$ and $p_{1}=\pm
e^{-iut^{\prime }}$, $p_{4}$ $=e^{iut^{\prime }}$. For positive case, the
atomic state is $|\Psi \rangle =\cos ut^{\prime }(|gg\rangle +|ee\rangle
)-i\sin ut^{\prime }(|eg\rangle +|ge\rangle )$ which is a always maximal
entanglement state. By choosing $ut^{\prime }=n\pi $ or $ut^{\prime }=n\frac{%
\pi }{2}$, one can obtain Bell state $|\Psi \rangle _{a}=\frac{1}{\sqrt{2}}%
(|ge\rangle +|eg\rangle )$ or $|\Psi \rangle _{a}=\frac{1}{\sqrt{2}}%
(|gg\rangle +|ee\rangle )$. Although during short time section, Concurrence
is not the maximum value. But it does not mean\ the entanglement can not be
completely transferred. Notice Fig.1 short time section, the amplitude of
the fields during this interval is small and the entanglement do not achieve
its maximum value. So, we can not retrieve a maximum entanglement if the
two-mode fields do not achieve its maximal entanglement. It is easy to see
from Eq.(15) that the probability of this projection is 25\%.

\subsection{ Other Projections $P=|\pm i\protect\lambda _{1}t^{\prime },i%
\protect\lambda _{2}t^{\prime }\rangle \langle \pm i\protect\lambda %
_{1}t^{\prime },i\protect\lambda _{2}t^{\prime }|$}

Except the projection of $P=|0,0\rangle \langle 0,0|$ , we still have other
six projections. If project the two-cavity fields with $P=|-i\lambda
_{1}t^{\prime },-i\lambda _{2}t^{\prime }\rangle \langle -i\lambda
_{1}t^{\prime },-i\lambda _{2}t^{\prime }|$ $($means we detect the cavity
fields as $|-i\lambda _{1}t^{\prime },-i\lambda _{2}t^{\prime }\rangle )$,
we have 
\begin{eqnarray}
p_{1} &=&e^{iut^{\prime }}\pm e^{-\frac{(\lambda _{1}t^{\prime })^{2}}{2}-%
\frac{(\lambda _{2}t^{\prime })^{2}}{2}-iut^{\prime }}, \\
p_{2} &=&e^{-\frac{(\lambda _{2}t^{\prime })^{2}}{2}+iut^{\prime }}\pm e^{-%
\frac{(\lambda _{1}t^{\prime })^{2}}{2}-2(\lambda _{2}t^{\prime
})^{2}-iut^{\prime }},  \nonumber \\
p_{3} &=&e^{-\frac{(\lambda _{1}t^{\prime })^{2}}{2}+iut^{\prime }}\pm e^{-%
\frac{(\lambda _{2}t^{\prime })^{2}}{2}-2(\lambda _{1}t^{\prime
})^{2}-iut^{\prime }},  \nonumber \\
p_{4} &=&e^{\frac{(\lambda _{1}t^{\prime })^{2}}{2}-\frac{(\lambda
_{2}t^{\prime })^{2}}{2}+iut^{\prime }}\pm e^{-2(\lambda _{1}t^{\prime
})^{2}-2(\lambda _{2}t^{\prime })^{2}-iut^{\prime }}.  \nonumber
\end{eqnarray}

\bigskip For this case, the entanglement transferred is shown in Fig. 4 line
a. During short evolution time, we can obtain a little entanglement. For
large time $t$, $p_{1}=e^{iut}$ and $p_{2}=p_{3}=p_{4}=0$. Thus, $|\Psi
\rangle _{a}=\frac{1}{2}(|g\rangle +|e\rangle )(|g\rangle +|e\rangle )$ is a
direct product state for positive case. We fail to transfer entanglement.
What we get is just to pumping the atomic state from $|g\rangle $ to
superposition $|g\rangle +|e\rangle )$ and obtain larger amplitude of the
field $i\lambda _{j}t^{\prime }$(first transformation the amplitude is ( $%
i\lambda _{j}t^{\prime }/2$). If project the two-mode fields with $%
P=|i\lambda _{1}t^{\prime },i\lambda _{2}t^{\prime }\rangle \langle i\lambda
_{1}t^{\prime },i\lambda _{2}t^{\prime }|$, the atomic entanglement behavior
is almost the same. 
\begin{figure}[h]
\includegraphics*[width=75mm,height=50mm]{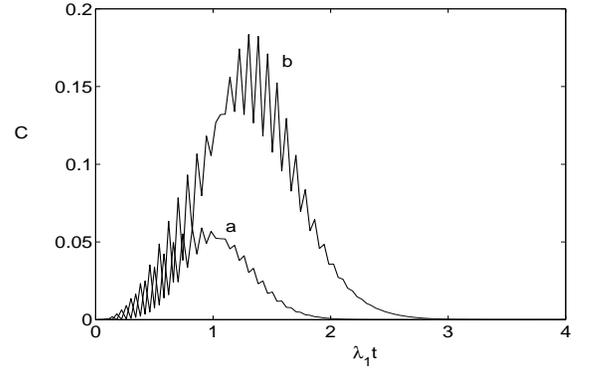}
\caption{The atomic entanglement change with time, corresponding to positive
in Eq.(12). Line a and b correspond to postselection $P=|-i\protect\lambda %
_{1}t^{\prime },-i\protect\lambda _{2}t^{\prime }\rangle \langle -i\protect%
\lambda _{1}t^{\prime },-i\protect\lambda _{2}t^{\prime }|$, $P=|-i\protect%
\lambda _{1}t^{\prime },0\rangle \langle -i\protect\lambda _{1}t^{\prime
},0| $, respectively.The parameters are the same with Fig. 2.}
\end{figure}

With the projection $P=|-i\lambda _{1}t^{\prime },0\rangle \langle -i\lambda
_{1}t^{\prime },0|$, we have%
\begin{eqnarray}
p_{1} &=&e^{-\frac{(\lambda _{2}t^{\prime })^{2}}{2}+iut^{\prime }}\pm e^{-%
\frac{(\lambda _{1}t^{\prime })^{2}}{2}-iut^{\prime }}, \\
p_{2} &=&e^{iut^{\prime }}\pm e^{-\frac{(\lambda _{1}t^{\prime })^{2}}{2}-%
\frac{(\lambda _{2}t^{\prime })^{2}}{2}-iut^{\prime }},  \nonumber \\
p_{3} &=&e^{-\frac{(\lambda _{1}t^{\prime })^{2}}{2}-\frac{(\lambda
_{2}t^{\prime })^{2}}{2}+iut^{\prime }}\pm e^{-2(\lambda _{1}t^{\prime
})^{2}-iut^{\prime }},  \nonumber \\
p_{4} &=&e^{-\frac{(\lambda _{1}t^{\prime })^{2}}{2}+iut^{\prime }}\pm
e^{-2(\lambda _{1}t^{\prime })^{2}-\frac{(\lambda _{2}t^{\prime })^{2}}{2}%
-iut^{\prime }}.  \nonumber
\end{eqnarray}%
The transferred entanglement is plotted in Fig.3 line b. At this case,
because the cavity mode 2 is projected with vacuum state (inverse the mode),
so the entanglement in short time evolution is larger than line a. But, for
large time $t$, $p_{2}=e^{iut}$ and $p_{1}=p_{3}=p_{4}=0$. The atomic state
is still a product one $|\Psi \rangle _{a}=\frac{1}{2}(|g\rangle +|e\rangle
)(|g\rangle -|e\rangle )$. Similarly, we can easy analyze the remaining
projection $P=|i\lambda _{1}t^{\prime },0\rangle \langle i\lambda
_{1}t^{\prime },0|$ and $P=|0,\pm i\lambda _{1}t^{\prime },0\rangle \langle
0,\pm i\lambda _{1}t^{\prime }|$. Their behaviors are like the case with the
projection $P=|-i\lambda _{1}t^{\prime },0\rangle \langle -i\lambda
_{1}t^{\prime },0|.$

Up to now, we have analyzed all kinds of projections. We conclude that with
the projection $P=|0,0\rangle \langle 0,0|$, we can completely transfer
two-cavity entangled coherent state into atomic state. \ We can understand
it from the reversible property of quantum mechanics. We produce entangled
coherent state from atomic entanglement and vacuum cavity mode. Projection
with $P=|0,0\rangle \langle 0,0|$ means keeping the two-mode still in vacuum
state; in this way the first process is reversed and entanglement will be
retrieved back into atomic state. Other projections can not make the
two-cavity state in its original one so that entanglement can not be
retrieved completely.

\section{Conclusion}

We have considered entanglement transfer between atomic state and two-mode
cavity state. By introducing two-mode classical fields, entanglement can be
transferred reciprocally. Before our proposal, people try to use
Jaynes-Cummings interaction to extract entanglement from continuous-variable
system. Due to the impossibility of perfect extraction, Serafini \cite%
{serafini} put forward a strategy by using multiple qubits (atomic cloud).
We, on the other hand, propose a scheme, by employing two classical fields,
which can perfect transfer between atomic state $|\Psi (0)\rangle _{12}=%
\frac{1}{\sqrt{2}}(|eg\rangle +|ge\rangle )$ and two-cavity entangled
coherent state $|\Psi (t)\rangle _{f\pm }=\frac{1}{\sqrt{M_{\pm }}}%
[e^{uti}|\alpha _{1},\alpha _{2}\rangle \pm e^{-uti}|-\alpha _{1},-\alpha
_{2}\rangle ]$. Because the entanglement measurement of two-cavity entangled
coherent state is very clear and easy, we do not face the hard problem of
quantifying entanglement of two-cavity non-Gaussian state. Our entanglement
measure quantifying on atomic state and two-cavity state is accurate. Our
scheme is also a good example on the transition between microscopic state
and macroscopic state.

Acknowledgments:This work was supported by Natural Science Foundation of
China under Grant No. 10575017 and Natural Science Foundation of Liaoning
Province of China under No. 20031073.

\end{document}